\documentstyle[12pt]{article}
\begin{document}
\newcommand{\newc}{\newcommand}
\newc{\ol}{\overline}
\newc{\ra}{\rightarrow}
\newc{\Xs}{cross section}
\newc{\gsim}{\buildrel{>}\over{\sim}}
\newc{\lsim}{\buildrel{<}\over{\sim}}
\newc{\gtau}{$\tau$}
\newc{\half}{\frac{1}{2}}
\newc{\beq}{\begin{equation}}
\newc{\eeq}{\end{equation}}
\newc{\barr}{\begin{eqnarray}}
\newc{\earr}{\end{eqnarray}}
\newc{\ua}{\uparrow}
\newc{\da}{\downarrow}
\title{Bell's Inequality and $\tau$-Physics at LEP}
\author{H.~Dreiner \\ Oct. 29$^{th}$, 1992 }
\date{{\small Theoretical Physics, University of Oxford,\\
1 Keble Road, Oxford OX1 3NP, England }}
\maketitle
\title{Abstract}
\begin{abstract}
\noindent In this talk presented at the TAU92 workshop, Colombus, OH, Sept. 92,
we summarize results presented in more detail in a recent
paper by S. Abel, M. Dittmar and the author where we gave a general proof that
Bell's inequality can not be tested at a collider experiment. In particular, a
measurement of correlated tau-spins at LEP does not constitute a test of local
realistic theories via Bell's inequality. The central point of the argument is
that such tests, where the spins of two particles are inferred from a
scattering
distribution, can always be described by a local hidden variable theory. In
response to questions at the workshop we go beyond the paper and show that an
old experiment involving the measurement of the correlated spins of the two
photons emitted in positronium decay via Compton-scattering is also not a
viable
test of Bell's inequality.

\end{abstract}

\section{Introduction:}
Since the 1935 paper \cite{EPR} of Einstein, Podolsky, and Rosen (EPR) the
question of whether quantum mechanics (QM) offers a complete description of
``physical reality" has received much attention. Since the work of Bell
\cite{bell1,bell2}, this question has been under the scrutiny of experiment.
Bell's theorem states that in all local-realistic theories \cite{belinf}
two-particle correlation functions satisfy Bell's inequality, whereas the
corresponding quantum mechanical correlation functions can violate the
inequality in some regions of parameter space. To date there are many
experimental tests of Bell's inequality \cite{exp1}-\cite{dm2} which we find
worth complementing in three respects \cite{us}:
\begin{itemize}
\item We think it would be of interest to test the inequalities with fermions
      and/or with massive particles; fermions  are peculiar to QM and a test
      with massive particles could exclude certain hidden-variable theories.
\item The correlation in the atomic cascade experiment is physically
established
      through electromagnetic interactions; there have been profound surprises
      in the physics of the weak-interactions, notably P- and CP-violation, and
      perhaps we are in store for another.
\item The energy scale probed in the atomic cascade experiments is in the $eV$
      range; as Bohm has pointed out, non-local effects may well become
apparent
      at a length-scale below about $10^{-13}\,cm$ \cite{bohm1}. It is thus
      important to test for non-local features at as high an energy-scale as
      possible.
\end{itemize}
In a recent paper \cite{us} we showed that it is {\it not} possible to test
locality via Bell's inequality in a collider experiment. It is therefore not
possible to devise a novel test involving the weak interactions at LEP for
example. We summarize the argument here.

\section{Proposed Experimental Test at LEP}
As a possible test of Bell's inequality with massive fermions we consider
$\tau$-production at LEP, $e^+e^-\ra Z^0\ra \tau^+\tau^-$. The two taus in the
$Z^0$-decay are produced with opposite chirality and thus with opposite
helicity ($M_{Z^0}\gg m_\tau$). Since the \gtau's have equal and opposite
momentum in the lab-frame, the \gtau-{spins} along their momentum axes are
{\it positively} correlated.

This system of two spin-$\half$ particles with positively correlated spins is
to
be considered for an experimental test of Bell's inequality. It has been shown
at LEP that the helicity of the tau can be determined with good accuracy from
the momentum spectrum of the decay products \cite{opal}. To simplify the
discussion we concentrate on the decay $\tau\ra\pi\nu_\tau$.

\medskip

The cross-section for the process $e^+e^-\ra \pi^+\pi^-{\bar\nu}_\tau\nu_\tau
$ calculated in the Standard Model is
\beq
\frac {d\sigma} {d\cos\theta_{\pi\pi}} (e^+e^-\ra
\pi^+\pi^-{\bar\nu}_\tau\nu_\tau)= A(1- \frac{1}{3} \cos\theta_{\pi\pi}),
\label{eq:crosssec}
\eeq
where the factor $A$ is given in \cite{us} and $\cos\theta_{\pi\pi}= ({\hat p}_
{\pi^+} \cdot {\hat p}_{\pi^-})$. The pion momenta $({\hat p}_{\pi^\pm})$ are
given in the respective parent \gtau\, rest-frames.

\section{Testing Bell's Inequality?}
In the Standard Model the expectation value for observing two pions at relative
angle $\cos \theta_{\pi\pi}$ is given by
\begin{eqnarray}
P_{QM}(\cos\theta_{\pi\pi})&=&\frac{
\frac {d\sigma}{d\cos\theta_{\pi\pi}} (e^+e^-\ra\pi^+\pi^-
\nu_\tau{\bar\nu}_\tau)} {\sigma(e^+e^-\ra\pi^+\pi^-\nu_\tau{\bar\nu}_\tau ) }
\\
&=& \half(1-\frac{1}{3}\cos\theta_{\pi\pi}).
\label{eq:qmexp}
\end{eqnarray}
A more general form of Bell's inequality than given in \cite{us}, which also
takes into account detector efficiencies is given in \cite{belinfeq}
\beq
2-| P(\cos\theta_{{\hat a}{\hat c}}) + P(\cos\theta_{{\hat a}{\hat d}}) |
 \geq | P(\cos\theta_{{\hat b}{\hat c}}) - P(\cos\theta_{{\hat b}{\hat d}}) |
\label{eq:bell}
\eeq
When inserting the form (\ref{eq:qmexp}) we obtain
\beq
12-| 6 - \cos\theta_{{\hat a}{\hat c}} - \cos\theta_{{\hat a}{\hat d}}| \geq
| \cos\theta_{{\hat b}{\hat c}} - \cos\theta_{{\hat b}{\hat d}} |,
\label{eq:bellexpl}
\eeq
which is satisfied for all sets angles $\theta_{{\hat {x}}{\hat y}}$. Here
$\theta_{{\hat {x}}{\hat y}}$ denotes the angle between two pions emitted in
directions $\hat x$ and $\hat y$ respectively. We thus conclude that as
presented this experiment is not a test of Bell's inequality.

It is worth pointing out that this conclusion does not directly depend on the
factor $\frac{1}{3}$ in Eq.(\ref{eq:qmexp}). If we consider a more general
form $P(\cos\theta)=\half(1-a\cos\theta)$, we find after insertion into
Eq.(\ref{eq:bell}) that it is satisfied for all $a\leq 1$:
\beq
\frac{4}{a} - | \frac{2}{a} - a\cos\theta_{{\hat a}{\hat c}} - a \cos
\theta_{{\hat a}{\hat d}}| \geq | \cos\theta_{{\hat b}{\hat c}} -
\cos\theta_{{\hat b}{\hat d}} |.
\eeq

\section{Conceptual Deficiency of the Experiment}
It is straightforward, using a construction proposed by Kasday in a
different context \cite{t6}, to construct an {\it ad hoc} local hidden variable
theory LHVT which also leads
to the result (\ref{eq:qmexp}) for the production of charged pions via $\tau
$-pairs at LEP. The construction goes as follows: the Standard Model (and thus
QM) provides the differential cross-section
which is a function of the unit pion momenta\footnote{ From $f({\hat p}_{\pi^
+}, {\hat p}_{\pi^-})$ we obtained Eq.(\ref{eq:crosssec}) by integrating over
all but the relative angle between the pions. The formula for $f( {\hat
p}_{\pi^
+}, {\hat p}_{\pi^-})$ is very extensive and we do not include it here. It
should however be clear that such a function exists and this is all that we
require.}
\beq
\frac{d\sigma}{d\Omega_+d\Omega_-} (e^+e^-\ra \pi^+\pi^-\nu_\tau{\bar
\nu}_\tau)
= f( {\hat p}_{\pi^+}, {\hat p}_{\pi^-}),
\label{eq:kasday1}
\eeq
where as before we consider the pion momenta in the respective rest frames of
the tau-leptons; and $d\Omega_\pm$ denotes the angular phase space of the two
pions respectively. Now let the hidden variables for each tau-lepton be a set
of
unit vectors $({\hat\lambda}_e, {\hat\lambda}_\mu, {\hat\lambda}_\pi,
{\hat\lambda }_\rho,\ldots)$. When the tau decays as
\makebox[1.0in]{$\tau^-\ra\pi^-\nu_\tau ,$} ${\hat\lambda }_\pi$ tells it to
decay such that the pion momentum is in the direction ${\hat p}_\pi=
{\hat\lambda}_\pi$. Now let $F( {\hat\lambda}_{ \pi^+}, {\hat\lambda}_{\pi^-})$
be the original probability distribution of the hidden variables. If we
identify
this LHVT with the QM function of Eq.(\ref{eq:kasday1})
\beq
F( {\hat \lambda}_{\pi^+}, {\hat \lambda}_{\pi^-}) = f( {\hat \lambda}_{\pi^+},
{\hat \lambda}_{\pi^-}),
\label{eq:kasday2}
\eeq
we obtain a hidden-variable theory which will exactly reproduce all
experimental
results of the process $e^+e^-\ra \pi^+\pi^-\nu_\tau{\bar \nu}_\tau$.

It is essential in the above construction that QM provides the function $
f( {\hat p}_{\pi^+}, {\hat p}_{\pi^-})$. This in turn is only possible because
all the components of ${\hat p}_{\pi^\pm}$ commute with each other
\begin{eqnarray}
 [ ({\hat{p}}_{\pi^+})_i , ({\hat{p}}_{\pi^-})_j ] & = & 0, \nonumber \\
 {[} ({\hat{p}}_{\pi^\pm})_i , ({\hat{p}}_{\pi^\pm})_j ] & = & 0 .
\end{eqnarray}
By contrast, when measuring non-commuting operators in the final state this
construction is no longer possible. For example, in the two photon experiments
\cite{t7}, when measuring the transverse spin components $S_a^i,$ $i=1,2,$
$ a=x,y,$ of the two correlated photons we have
\beq
[ S_{a}^1,S_{b}^2 ]  =  0, \qquad {[} S_{a}^{i},S_{b}^{i}] =
i\hbar \epsilon_{abz}S^i_z.
\eeq
In this case QM does {\it not} provide a function $f(S_x^i,S_y^i,...)$ because
QM can not give a function for an experimental result which depends on the
simultaneous knowledge (measurement) of two non-commuting variables. Therefore
the Kasday construction breaks down.

We conclude that in all experiments where the correlated observables commute,
it
is possible using the method of Eq.(\ref{eq:kasday2}) to construct a LHVT which
reproduces the QM results and thus the experimental results. In these cases the
QM result must satisfy Bell's inequality just like the LHVT result does. In the
case of an alternative experiment which measures non-commuting observables this
is not the case and we are left with the {\it possibility} that it is a viable
test of Bell's inequality. Thus non-commuting observables are a {\it necessary}
requirement for a valid test of locality.

Our conclusion similarly applies to the proton fixed target experiment
\cite{t8}; the $\Lambda$ pair production experiment \cite{t1,dm2} and the
positronium decay experiments \cite{positronium}, which we return to below.

\section{Testing Hidden Variable Theories}
Despite the above criticism several experiments involving commuting observables
have been performed \cite{positronium}-\cite{dm2}. We briefly outline the
method
employed and the assumptions made in order to extract a correlation function
from the experiments which does violate Bell's inequality. We go on to outline
why we disagree with this method.

Given the correlation $P_{\pi\pi}(\theta) = \half(1-\frac{1}{3}\cos
\theta_{\pi\pi})$, it is clear if we consider the function
\beq
P(\theta)= 3(P_{\pi\pi} (\theta) -\half)=-\cos\theta
\label{eq:theta}
\eeq
we find that for special sets of angles $P(\theta)$ violates Bell's inequality
(\ref{eq:bell}). Presented in this form $P(\theta)$ has no physical
interpretation and it is thus irrelevant whether it violates Bell's inequality
or not. Recall that the objective of experimental tests of Bell's inequality is
to test whether nature is described by a local theory or by a non-local theory
such as QM. We now proceed to show that {\it if} one assumes that QM is the
proper description (despite the fact that we want to test it!) {\it then}
$P(\theta)$ can be given a physical interpretation and given this strong
assumption one can ``test Bell's inequality".

Within the framework of QM one can derive \cite{us} on symmetry grounds a
general formula for the expectation value of the relative angle between the
final state pions emitted from the spin-$\half$ \gtau's
\beq
P_{\pi\pi}(\theta) = c_1+c_2 \,\,< ({\hat p}_{\pi^+}\cdot
{\vec\sigma}_{\tau^+})( {\hat p}_{\pi^-}\cdot {\vec\sigma}_{\tau^-})>_{QM} .
\label{eq:pqft2}
\eeq
for general constants $c_1$ and $c_2$. This corresponds to $s$- and $p$-wave
scattering. The second term on the right-hand side is the expectation value for
the $\tau^+$-spin to be observed in the direction ${\hat p}_{\pi^+}$ {\it and}
the $\tau^-$-spin in the direction ${\hat p}_{\pi^-}$. But keep in mind that
this is all within the framework of QM. This second term has an exact analogon
in the better known atomic cascade experiments \cite{t7}, where the transverse
spins of two photons, $\gamma_a,\gamma_b$, in the singlet state are measured by
standard polarizers set-up in (unit-vector) directions $\hat a$ and $\hat b$.
The expectation value that the two photons are observed is then given by
\begin{eqnarray}
P_{\gamma_a\gamma_b} &=& < ({\hat a}\cdot {\vec\sigma}_{\gamma_a})
({\hat b}\cdot {\vec\sigma}_{\gamma_b})  >_{QM}, \\
&=& -{\hat a}\cdot {\hat b}.
\end{eqnarray}
This expectation value violates Bell's inequality for certain sets of angles.
We
can see from Eq.(\ref{eq:pqft2}) that within the framework of QM the
expectation
value of the correlation of the \gtau-spins
\beq
P_{\sigma_\tau\sigma_\tau}(\theta)= \,\, <({\hat a}\cdot{\vec
\sigma}_{\tau^+})({\hat b}\cdot {\vec\sigma}_{\tau^-})>_{QM}
\eeq
can be given in terms of the expectation value of the correlated pion momenta
\beq
{ P}_{\sigma_\tau\sigma_\tau}(\theta)= \frac{P_{\pi\pi}(\theta)
-c_1}{c_2}=-\cos
(\theta),
\label{eq:ptau}
\eeq
which is just $P(\theta)$ of Eq.(\ref{eq:theta}). If we now make the in our
opinion strong assumption that this is true in general, {\it independently} of
whether QM describes the correlation correctly or not (remember that is what we
are trying to check) and eventhough we used QM to obtain Eq.(\ref{eq:ptau}),
then we can experimentally test $P_{\sigma_\tau\sigma_\tau }(\theta)$ and the
correlation of the tau spins by measuring $P_{\pi\pi}(\theta) $. As given
above,
$P_{\sigma_\tau \sigma_\tau}(\theta)$ violates Bell's inequality and is thus
not
reproduceable by any LHVT.

It is tempting to conclude that this is a test of locality via Bell's
inequality. We argue instead that it is merely a test of a subclass of LHVTs,
where the size of the subclass is unknown. In obtaining Eq.(\ref{eq:ptau}) we
have used QM to translate our expectation of the pions ($P_{ \pi\pi}$) into an
expectation about the tau-spins ($P_{\tau\tau} $). Thus we are assuming QM to
start with. Dynamically the connection is given through the tau-decay and we
are
therefore specifically assuming QM ({\it i.e.} the SM) describes the tau-decay
properly {\it as well as} the correlations we establish or maintain through it.
However, our main objective is to test whether QM offers a complete description
of nature, and thus it is inconsistent to employ it in the test. If we do make
the assumption Eq.(\ref{eq:ptau}) or Eq.(\ref{eq:pqft2}) then of course we are
restricting ourselves to a test of only those LHVTs which also satisfy
Eq.(\ref{eq:pqft2}). Thus we are testing a subclass of LHVTs, but there is a
subclass of LHVTs of unknown size which remains untested. The LHVT we
constructed in section 4 for example falls into the second category.

One might now reply that we know from measurements on single taus that QM
describes the tau decay correctly. Indeed we do, but it is also known that it
is
trivial to construct a LHVT which correctly describes the decay of a single
tau as well. The distinction between na\"{\i}ve local-realistic expectations
(LHVTs) and QM doesn't arise until we consider an experiment involving
the correlation of at least two particles. This point is at the heart of the
EPR
paradox as well as the work of Bell. Thus it is exactly at the level of the
established correlation that we want to test QM {\it versus} LHVTs and it is
not appropriate to assume the correctness of QM in order to consider an {\it
unobserved} correlation.

The above argument might be considered to be implicit in Kasday's paper
\cite{t6}. However, in the subsequent literature it has been ignored
\cite{t8,t1,dm2,wheeler,privitera}. We have thus found it important to stress
and elucidate the full extent of the argument.

\section{Compton Scattering Experiment}
In response to discussions at the workshop we conclude with a brief review of
an
old experimental test of Bell's inequality where polarization correlations are
observed for pairs of photons emitted in the decay of positronium
\cite{positronium}. From parity considerations it is known that the photons are
in the singlet state. The spins of the $500\,keV$ photons are ``analyzed" via
Compton scattering, and the correlation is thus an angular correlation in the
momenta of the scattered photons. This is thus an experiment where commuting
observables are measured and from our proof we conclude that the experiment is
{\it not a test of Bell's inequality}.

This can also be seen explicitly from the QM formula for the scattering
correlation. The correlation is given by the Klein-Nishina Compton scattering
formula as for example given in \cite{belinf}
\beq
P(\theta_1,\phi_1,\theta_2,\phi_2)= \frac{1}{16\pi^2} (1- C
\frac { G(\theta_1) G(\theta_2)} { F(\theta_1) F(\theta_2)} \cos
2(\phi_1-\phi_2))
\label{eq:klein}
\eeq
where we have introduced a normalization factor as in Eq.(2) and
\barr
F(\theta) &=& [2 + (1-\cos\theta)^3]/(2-\cos\theta)^3 ,
\\
G(\theta) &=& \sin^2\theta/(2-\cos\theta)^2.
\earr
The constant $C\leq \half$ and $\theta_i,\phi_i$ are the final state angles
of the photons $i=1,2$. As an example consider
$P(\theta_1,\phi_1,\theta_2,\phi_2)$ as a function only of the relative angle
$\Delta\phi=\phi_1-\phi_2$ and with fixed $\theta_1,\theta_2$
\beq
P(\Delta\phi) = \frac{1}{\pi} (1-
C \,\frac { G(\theta_1) G(\theta_2)} {F(\theta_1) F(\theta_2) } \cos
2(\Delta\phi)),
\eeq
It is straightforward to show that $G(\theta)/F(\theta)<1$ for all $\theta$ and
thus using the argument of Eq.(6) Section 3 we see that $P(\Delta\phi)$
satisfies Bell's inequality for any angles $\Delta\phi,\theta_1,\theta_2$.

We also point out, that Bell constructed a LHVT \cite{lhvtbell} which gives
{\it
exactly} the same prediction for the correlation of the two scattered photons
as
the QM prediction given in Eq.(\ref{eq:klein}). The statement of Bell's theorem
\cite{bell1} is that {\it every} LHVT must satisfy Bell's inequality. Thus the
LHVT constructed by Bell and the identical QM prediction given in
Eq.(\ref{eq:klein}) must satisfy Bell's inequality for {\it all} angles $\theta
_1,\phi_1,\theta_2, \phi_2$. And we conclude that this experiment is not a
viable test of Bell's inequality.

\medskip

\noindent {\bf Acknowledgements} The author would like to thank Steven Abel
and Michael Dittmar with whom this work was done.

\end{document}